\documentclass[12pt]{article}
\usepackage[textwidth=6.25in]{geometry}

\usepackage{amsmath} 
\usepackage{times}
\usepackage{amsmath,amsthm,amssymb}
\usepackage{enumerate}
\usepackage{graphicx}
\usepackage{natbib}
 \usepackage{setspace}
\usepackage{subcaption}
\usepackage{color}
\usepackage{sectsty}
\usepackage{titlesec}
\titleformat{\section}[hang]{\centering \bfseries\large}{\thesection.}{0.4em}{}

\renewcommand{\P}{\mathbb{P}}
\newcommand{\E}{\mathbb{E}}
\newcommand{\COR}{\text{COR}}

\newtheorem{theorem}{Theorem}[section]

\newtheorem{proposition}[theorem]{Proposition}

\theoremstyle{definition}
\newtheorem{example}[theorem]{Example}
\theoremstyle{definition}

\def\pb{\overline{p}}

\def\one{{\bf 1}}
\def\v{{\bf v}}
\def\F{{\cal F}}
\def\G{{\cal G}}
\def\P{{\mathbb P}}
\def\E{{\mathbb E}}

\def\Cov{{\rm Cov}\,}
\def\ee{\varepsilon}
\def\|{\, | \,}
\def\probit{p_{\rm probit}}

\def\plog{p_{\rm log}}
\def\conv{{\rm conv}}

\title{\vspace{-0em} \Large Modeling Probability Forecasts via Information Diversity}
\author{\vspace{-1em}Ville A. Satop\"a\"a, Robin Pemantle, and Lyle H. Ungar\thanks{Ville A. Satop\"a\"a is a Doctoral Candidate, Department of Statistics, The Wharton School of the University of Pennsylvania, Philadelphia, PA 19104-6340 (e-mail: satopaa@wharton.upenn.edu); Robin Pemantle is a Mathematician, Department of Mathematics, University of Pennsylvania, Philadelphia, PA 19104-6395 (e-mail: pemantle@math.upenn.edu); Lyle H. Ungar is a Computer Scientist, Department of Computer and Information Science, University of Pennsylvania, Philadelphia, PA 19104-6309 (e-mail: ungar@cis.upenn.edu). This research was supported in part by NSF grant \# DMS-1209117 and a research contract to the University
of Pennsylvania and the University of California from the Intelligence
Advanced Research Projects Activity (IARPA) via the Department of
Interior National Business Center contract number D11PC20061. The
U.S. Government is authorized to reproduce and distribute reprints for
Government purposes notwithstanding any copyright annotation
thereon. Disclaimer: The views and conclusions expressed herein are
those of the authors and should not be interpreted as necessarily
representing the official policies or endorsements, either expressed
or implied, of IARPA, DoI/NBC, or the U.S. Government.
The authors would like to thank Edward George and Shane Jensen for helpful discussions.}}
\date{\vspace{-8.5ex}}
\sectionfont{\centering}

\begin{document}
\maketitle
%\pagestyle{myheadings}
%\markboth{Understanding Probability Extremizing}{Satop\"a\"a et al.}
\begin{abstract}
\singlespace
Randomness in scientific estimation is generally 
assumed to arise from unmeasured or uncontrolled factors. However, 
when combining subjective probability estimates, heterogeneity
stemming from people's cognitive or information diversity is often
more important than measurement noise.  This paper presents a novel
framework that uses partially
overlapping information sources.  A specific model is proposed
within that framework and applied to the task of aggregating
%that models the heterogeneity arising from forecasters that use 
%partially overlapping information sources, and applies that model to 
%the task of  aggregating
 the probabilities given by a group of forecasters 
who predict whether an event will occur or not. Our model describes 
the distribution of information across forecasters in terms of easily
interpretable parameters and shows how the optimal amount
of \textit{extremizing} of the average probability forecast (shifting
it closer to its nearest extreme) varies as a function of the forecasters'
information overlap.  Our model thus gives a more principled
understanding of the historically {\it ad hoc} practice of extremizing
average forecasts. Supplementary material for this article is available
online.
\end{abstract}

\newpage
\section{INTRODUCTION AND OVERVIEW}
%\end{center}

\subsection{The Forecast Aggregation Problem\nopunct}
Probability forecasting is the science of giving probability estimates for future events.
%, such as possible rain tomorrow or  
%% For instance, the department of foreign affairs may be interested in
%whether war will break out in Egypt
%this year.
%Even though the classical examples are meteorological
%\citep{murphy1977reliability}, predicting, for example, whether rain will occur in a given time period, the practice has 
%spread to many other fields as well. This includes medical diagnosis  \citep{pepe2003statistical}, estimation of credit default
%\citep{kramer2006evaluating}, and predicting geopolitical events such as who will win
%the next Congolese election or whether war will break out in Egypt
%this year \citep{tetlock2005expert}. 
%Accurate forecasts on events like these are particularly important to the decision-maker, who must decide, for instance, what to pack
%for tomorrow's family outing or whether to advise against all travel
%to Egypt.
Typically more than one different forecast is available on the same event. Instead of trying to guess which prediction is the most accurate, the predictions should be combined into a single consensus forecast \citep{armstrong2}. Unfortunately, the forecasts can be combined in many different ways, and the choice of the combination
rule can largely determine the predictive quality of the final
aggregate.  
%Analyzing and developing improved rules is the main motivation of 
This is the principal motivation for the problem of  \textit{forecast aggregation} that aims to combine multiple forecasts into a single
forecast with optimal properties.

There are two general approaches to forecast aggregation: empirical
and theoretical. 
% The empirical approach is akin to machine
%learning. 
Given a training set with multiple forecasts on events with
known outcomes, the empirical approach experiments with different
aggregation techniques and chooses the one that yields the best performance on the training set.
The theoretical approach, on the other hand, first constructs a
probability model and then computes the optimal aggregation procedure
under the model assumptions.  
%This may involve estimating model
%parameters from the forecasts. 
Both approaches are important.  Theory-based procedures that do not
perform well in practice are ultimately of limited use.  On the other
hand, an empirical approach without theoretical underpinnings lacks
both credibility (why should we believe it?)  and guidance (in which
direction can we look for improvement?). As will be discussed below, the history of forecast aggregation to date is largely
empirical. 
%(Not only the approaches in
%Section~\ref{ss:empirical} but also the approaches described in
%Section~\ref{ss:measurement}, to which we have provided theoretical
%underpinnings, have been pursued in a manner that is largely
%empirical.)

The main contribution of this paper is a plausible
theoretical framework for forecast aggregation called the \textit{partial information framework}. Under this framework, forecast heterogeneity stems from information available to the forecasters and how they decide to use it. For instance, forecasters studying the same (or different) articles on the presidential election may use distinct parts of the information and hence report different predictions of a candidate winning. 
%one forecast might depend on television news in the United States
%while the other is based on local news in Egypt. 
%The framework allows for
%the possibility that the forecasters have the same information but use it differently. 
Second, the framework allows us to interpret existing aggregators and illuminate aspects
that can be improved. This paper specifically aims to clarify the practice of {\em probability
extremizing}, i.e., shifting an average aggregate closer to its nearest
extreme. Extremizing is an empirical technique that has been widely used to
improve the predictive performance of many simple aggregators such as
the average probability. Lastly, the framework is applied to a
specific model under which the optimal aggregator can be computed.

\subsection{Bias, Noise, and Forecast Assessment}
\label{BiasNoise}
Consider an event $A$ and an indicator function $\one_A$ that equals one or zero depending whether $A$ happens or not, respectively. There are two common yet philosophically different approaches to linking $A$ with the probability forecasts. The first assumes $\one_A \sim \text{Bernoulli}(\theta)$, where $\theta$ is deemed a ``true'' or ``objective'' probability for $A$, and then treats a probability forecast $p$ as an estimator of $\theta$ (see, e.g., \citealt{lai2011evaluating}, and Section \ref{ss:measurement} for further discussion). The second approach, on the other hand, treats $p$ as an estimator of $\one_A$. This links the observables directly and avoids the controversial concept of a ``true'' probability; for this reason it is the approach adopted in this paper. 

As is the case with
all estimators, the forecast's deviation from the truth can be broken into bias and noise. Given that these components are typically handled by different mechanisms, it is important, on the theoretical level, to consider them as two separate problems. 
%On the
%theoretical level, it is important to separate the problems of bias
%and noise because they are solved by different mechanisms. 
This paper focuses on noise reduction. Therefore, each forecaster is considered 
% In the forecasting literature, a forecaster
% 
%is called \
\textit{calibrated}. Here calibration is defined in terms of conditional expectation and 
hence represents a property of  
%where the expectation is computed with respect to the conditional distribution of $\one_A$ given $p$. 
%Therefore this form of calibration is a property of 
the underlying joint distribution of  $\one_A$ and $p$. More specifically, the forecast $p$ is calibrated for the outcome $\one_A$ if $\P(\one_A = 1|p) = \E(\one_A | p) = p$ almost surely.
% This is a conditional expectation and hence represents a property of   
%%where the expectation is computed with respect to the conditional distribution of $\one_A$ given $p$. 
%%Therefore this form of calibration is a property of 
%the underlying joint distribution of  $\one_A$ and $p$. 
This form of calibration was alluded to by \citet{murphy1987general} and mentioned possibly even earlier than that. Over the years it has become common in the statistical and meteorological forecasting literature (see, e.g., \citealt{Ranjan08}; \citealt[Section 7.2.2.]{jolliffe2012forecast} for recent references). It is, however, different from  the notion of empirical calibration discussed by \citet{dawid1982well}, \citet{foster1998asymptotic}, and many others.

A forecast (individual or aggregate) is typically assessed with a loss
function $L(p , \one_A)$.  A loss function is called 
{\em proper} or {\em revealing} if the Bayesian optimal strategy is
to tell the truth.  In other words, if the subjective probability
estimate is $p$, then $t = p$ should minimize the expected loss $p
L(t,1) + (1-p) L(t,0)$.   Therefore, if a group of
sophisticated forecasters operates under a proper loss function,
the assumption of calibrated forecasts is, to some degree,
self-fulfilling. There are, however, many different proper loss functions, and an estimator that outperforms another under one loss function will not necessarily do so under a different one. For example, minimizing the quadratic loss function $(p - \one_A)^2$, also known as the Brier score, gives the estimator with the least variance.
%The most common proper loss function is the quadratic loss $(p - \one_A)^2$, also known as
%the {\em Brier score}. This gives the estimator with the least variance. 
%Many other choices, however, are possible, and
%an estimator that outperforms another under one loss function will not
%necessarily do so under a different one. 
This paper concentrates on minimizing the variance of the aggregators, though
 much of the discussion holds under general proper loss
functions. See  \cite{HwPe1997}
for a discussion of proper loss functions.
%
%The point of this brief review is two-fold.  First, when a group of
%sophisticated forecasters operates under a proper loss function,
%the assumption of calibrated forecasts is, to some degree,
%self-fulfilling. Secondly, the
%assessment of the aggregated forecast is not uni-dimensional any more
%than is the assessment of an individual forecast.  In order to compare
%different aggregation procedures, a scoring rule must be chosen.  In
%practice, the Brier score is the most common choice, perhaps because
%of its simplicity or because of the
%paramount status of variance in the statistical literature. This paper concentrates on minimizing the variance of the aggregators, though
% much of the discussion holds under general proper loss
%functions.

\subsection{The Partial Information Framework}
\label{PIFintro}
%As is the case with any probability model for forecast aggregation, t
The construction  of the partial information framework begins with
%Consider
 a probability space $(\Omega, \F , \P)$ and a
measurable event $A \in \F$ to be forecasted by $N$ forecasters. These forecasters operate under the same probability model but make predictions based on different information sets. More specifically, in any Bayesian setup,
with a proper loss function, it is more or less tautological that
Forecaster $i$ reports $p_i := \E (\one_A \| \F_i)$, where $\F_i \subseteq \F$ is
the \textit{information set} used by the forecaster. Therefore $\F_i \neq \F_j$ if $p_i \neq p_j$, and forecast heterogeneity stems purely from \textit{information diversity}.  Note, however, that if Forecaster $i$ uses a simple rule, $\F_i$ may not be the full $\sigma$-field of information available to the forecaster but rather a smaller $\sigma$-field corresponding to the information used by the rule. For example, when forecasting the re-election of the president,  a forecaster obeying the dictum ``it's the economy,
stupid!''  might utilize a $\sigma$-field containing only economic
indicators.  Furthermore, if two forecasters have access to the same $\sigma$-field, they may decide to use different sub-$\sigma$-fields, leading to different predictions. 
%This is reminiscent of a forecasting algorithm that only uses
%a similarly restricted subset of information. 
Therefore,
information diversity does not only arise from differences in the available information, but also from how the forecasters decide to use it.

The person performing the aggregation is assumed to know only $\F_0 = \{\emptyset, \Omega\}$, namely the trivial $\sigma$-field. Given that every forecaster knows at least as much, the aggregator readily adopts any forecaster's prediction without modification. Therefore, each forecaster is considered to be an ``expert'' in the sense introduced in \cite{degroot1988bayesian} and later discussed in \cite{dawid1995coherent}.

%The probability measure $\P$ provides a prior probability on $A$. Given that $\P(A) = \E(\one_A | \F_0)$, where $\F_0 = \{\emptyset, \Omega\}$ is the trivial $\sigma$-field, and $\F_0 \subseteq \F_i$ for all $i = 1, \dots, N$, this prior is based on information considered known to all the forecasters and the aggregator. If $\F_i \neq \{\emptyset, \Omega\}$, the forecaster is assumed to have additional information.  Given that the aggregator only knows $\F_0$, it is 

%Each forecaster, however, may also have additional information and is therefore considered to be an ``expert'' in the sense discussed in \cite{dawid1995coherent}.
%

Under the partial information framework the forecasts are calibrated.  This can be verified by direct computation as follows:
\begin{align*}
\E(\one_A | p_i) &= \E\{\E(\one_A | p_i, \F_i) | p_i\} = \E\{\E(\one_A | \F_i) | p_i\} = \E(p_i | p_i) = p_i.
\end{align*}
Conversely, if $p_i$ is any calibrated forecast, then $p_i = \E(\one_A | \G_i)$, where $\G_i = \sigma(p_i) \subseteq \F_i$ is the $\sigma$-field generated by $p_i$. This shows constructively that assuming the general form $p_i = \E (\one_A \| \F_i)$ does not pose any additional restrictions but arises directly from the assumption of calibration and the existence of an underlying probability model. The $\sigma$-field $\G_i$, however, corresponds to the information revealed by the forecast and hence may not be equal to the full $\sigma$-field of information actually used by the forecaster, namely $\F_i$.

The distinction between $\F_i$ and $\G_i$ introduces
%The partial information framework distinguishes
 two benchmarks for aggregation efficiency. 
% They differ in the degree to which they 
The first is the {\em oracular} aggregator
$p' := \E (\one_A \| \F')$, where $\F'$ is the $\sigma$-field
generated by the union of the information sets $\{\F_i : i = 1, \dots, N\}$. This field represents all the information used by the forecasters. Given that aggregation cannot be improved beyond using all the information  of the forecasters, the oracular aggregator represents a theoretical optimum and is therefore a reasonable upper bound on estimation efficiency. 
%Note that if a forecaster chooses to discard
%some of the available information, then, for the purposes of aggregation,
%that information may as well not exist. 

In practice, however, information comes to the
aggregator only through the forecasts $\{p_i : i = 1, \dots, N\}$. Given that $\F'$ generally cannot be constructed from these forecasts alone, no practically feasible aggregator can be expected to perform as well as $p'$.  Therefore, a more achievable benchmark is the \textit{revealed} aggregator $p'' := \E (\one_A \|
\F'')$, where $\F''$ is the $\sigma$-field generated (or revealed) by the forecasts
$\{ p_i : i = 1, \dots, N \}$, or equivalently by the union of the generated $\sigma$-fields $\{\G_i : i = 1, \dots, N\}$.

%This benchmark minimizes
%the expectation of any proper loss function \citep{Ranjan08} and can be potentially applied in practice. 

%For this reason it is the relevant estimator in each specific partial information model. 

%Even though the partial information framework, as specified above, is too theoretical for direct application, it highlights the crucial components of information aggregation and hence facilitates the design of a closely-related abstraction. This paper develops such an abstraction and calls it the Gaussian partial information model. Under this model, the information among the forecasters is summarized with a particular covariance structure, which allows for a flexible construction of many application specific aggregators. 

Even though the partial information framework, as specified
        above, is too theoretical for direct application, it highlights
        the crucial components of information aggregation and hence
        facilitates formulation of more specific models within the
        framework.  This paper develops such a model and calls it the
        Gaussian partial information model.  Under this model, the
        information among the forecasters is summarized by a covariance
        structure. This provides sufficient flexibility to allow
        for construction of many application-specific aggregators.

% The model is rather general and can be easily adapted to different applications by changing the assumptions about the structure of the forecasters' information. 

%The final section of the paper investigates one such assumption, namely symmetric information. 

%Both of these benchmarks and also the individual forecasts are conditional expectations of $\one_A$ given different sub-$\sigma$-fields. Such conditional estimators are always calibrated. In particular, the revealed aggregator is calibrated if the individual forecasts are calibrated.  This should be viewed as a minimum criterion for noise-reducing aggregation. Aggregators that satisfy this property differ and can be evaluated on their ability to combine information from the individual sources. An aggregator that efficiently combines this information typically produces an estimate that is closer to $\one_A$ and hence has a lower proper loss.  

%Therefore the framework responds to \citet{Ranjan08} who proved that all convex combinations of calibrated forecasts are necessarily uncalibrated. This result invalidated theoretical justification of many commonly used aggregators and urged a .

%The {\em general partial information model} 
%is precisely the model described in the foregoing paragaphs.  
%Specific partial information models involve assumptions on
%the structure of $\{ \F_i \}$.  In each case, the relevant estimator 
%is the revealed aggregator. 

\subsection{Organization of the Paper}

The next section reviews prior work 
%that has focused largely on two aggregation frameworks which differ in their assumptions about the source of forecast heterogeneity. 
on forecast aggregation and relates it to the partial information framework.
Section~\ref{sec:model} discusses illuminating examples and
motivates the Gaussian partial information model.
% that is used in the remainder of the paper.
%our subsequent choices of specific partial information
%models.  
Section~\ref{extremizing} compares the oracular aggregator with the average probit score, thereby explaining the
empirical practice of probability extremizing.
Section~\ref{aggregation} derives the revealed aggregator and evaluates one of its sub-cases on real-world
forecasting data.  The final section concludes with a summary and discussion of
future research.

\section{PRIOR WORK ON AGGREGATION}
\label{sec:prior}
%Prior work on forecast aggregation has focused largely on two frameworks that differ in their assumptions about the source of forecast heterogeneity.  These approaches are discussed in the next subsections.

\subsection{The Interpreted Signal Framework}
\label{ss:inerpreted}

\citet{hong2009interpreted} introduce the {\em interpreted signal
framework} in which the forecaster's prediction is based on a personal
interpretation of (a subset of) the factors or cues that influence the
future event to be predicted.  Differences among the 
predictions are ascribed to differing interpretation procedures.  For
example, if two forecasters follow the same political campaign speech,
one forecaster may focus on the content of the speech while the other may
concentrate largely on the audience interaction.  Even though the
forecasters receive the same information, they interpret it
differently and therefore are likely to report different estimates of the probability that the candidate wins the election. Therefore forecast heterogeneity is assumed to stem from ``cognitive
diversity''.  

This is a very reasonable assumption that has been analyzed and
utilized in many other settings.  For example,
~\citet{parunak2013characterizing} demonstrate that optimal
aggregation of interpreted forecasts is not constrained to the
convex hull of the forecasts; \citet{broomell2009experts} analyze
inter-forecaster correlation under the assumption that the cues can be
mapped to the individual forecasts via different linear regression
functions.
% \citet{degroot1991optimal} analyze the linear opinion pool under a setup where  each forecaster first observes a random quantity $X_i$ and then reports $\P(A | X_i)$. 
%They  acknowledge that without a distributional model on the $X_i$'s the optimal aggregator cannot be formed. Instead of constructing a plausible model, they proceed to analyze a particular form of aggregation, namely the linear opinion pool. 
%Any linear opinion pool of calibrated forecasts, however, is today known to be both uncalibrated and under-confident \citep{Ranjan08}.
%Some preliminary steps have been made towards a more concrete model by, for example, assuming
%conditional independence of the $x_i$, \cite{}, but
%lacking a distributional model. 
To the best of our knowledge, no previous work has discussed a formal framework that explicitly links the interpreted forecasts to their target quantity. Consequently, the interpreted signal framework, as proposed, has remained relatively abstract. 
%The interpreted signal framework, when formalized, more or less implies the partial information framework.
%The present paper, however, formalizes the intuition behind it. 
%The interpreted signal framework, when formalized, more or less implies the partial information framework. 
The partial information framework, however, formalizes the intuition behind it and permits models with quantitative predictions.
%and provides a flexible construction that can be adopted to a broad range of forecasting setups. 

%
%The interpreted signal framework, when formalized, more or less
%implies the partial information framework.  Unfortunately, previous
%work on interpreted forecasts has produced only abstract concepts and
%has not been taken to the level of a precise formal model with
%quantitative predictions.  Given this lack of a quantitative model, in
%practice it is often easier and more common to explain subjective
%forecasts with the {\em measurement error framework}. This framework
%is described in the next subsection.

\subsection{The Measurement Error Framework}
\label{ss:measurement}
In the absence of a quantitative
interpreted signal model, prior applications have typically relied on
%explained forecast heterogeneity  with
% standard statistical models. These models are 
%different formalizations of 
 the \textit{measurement error framework} that generates forecast heterogeneity from a probability distribution. More specifically, the framework assumes a ``true'' probability
$\theta$, interpreted as the forecast made by an ideal forecaster, for the event $A$. The forecasters then ``measure'' some transformation of this probability $\phi(\theta)$ with mean-zero idiosyncratic error. Therefore each forecast is an independent draw from a common
probability distribution centered at $\phi(\theta)$, 
%, and the underlying probability model is just the classical statistical model for measurement with i.i.d. mean zero error. Consequently, 
and a recipe for an aggregate forecast is given by the average 
\begin{align}
\phi^{-1} \left\{ \frac{1}{N} \sum_{i=1}^N \phi
   (p_i) \right \}.
\end{align}
Common choices of $\phi(p)$ are the identity $\phi(p) = p$, the log-odds $\phi(p) = \log\left\{p/(1-p)\right\}$, and the probit $\phi(p) = \Phi^{-1}(p)$, giving three aggregators denoted in this paper with $\pb$, $\plog$, and $\probit$, respectively. These \textit{averaging aggregators} represents the main 
%Such \textit{averaging aggregators} illustrate 
 advantage of the measurement
error framework: simplicity.
%  The formula for a new aggregator is
%obtained in a straightforward manner by transforming, averaging, and
%transforming back.  
%The underlying probability model is also simple
%and very familiar.  

Unfortunately, there are a number of disadvantages. 
%First, the introduction of a nonlinear transformation biases the
%estimator.  Consequently, $\plog$ nor $\probit$ produces a
%calibrated forecast. 
%Exactly how problematic this is depends on
%whether the bias is small relative to noise or other
%adjustments. 
First, given that the averaging aggregators target $\phi(\theta)$ instead of  $\one_A$, 
%given that a measurement error model does not incorporate an explicit link between the forecasts and the correct quantity of interest, namely $\one_A$, 
important properties such as calibration cannot be expected. In fact, the averaging aggregators are uncalibrated and under-confident, i.e., too close to $1/2$, even if the individual forecasts are calibrated \citep{Ranjan08}.

%In some cases,  the direction of the bias can be determined. For instance, any convex
%combination of calibrated forecasts is known to be both uncalibrated and under-confident, i.e., too close to $1/2$ \citep{Ranjan08}.
%, i.e., 
%too close to $1/2$. 
%This applies to the averaging aggregators that are by nature different convex combinations of the
%forecasts.

Second,  the
underlying model is rather implausible. Relying on a true probability $\theta$ is vulnerable to
many philosophical debates, and even if one eventually manages to
convince one's self of the existence of such a quantity, it is
difficult to believe that the forecasters are actually seeing $\phi(\theta)$  with independent noise. Therefore, whereas the
interpreted signal framework proposes a micro-level explanation, the
measurement error model does not; at best, it forces us to imagine that
the forecasters are all in principle trying to apply the same
procedures to the same data but are making numerous small mistakes. 
%\textcolor{red}{If you do not believe the model can you truly believee that the aggregate represents consenus belief; this is somewhat different from predicting well.}

Third, the averaging aggregators do not often perform
very well in practice. For one thing,  \citet{hong2009interpreted} demonstrate that the standard
assumption of conditional independence poses an
unrealistic structure on interpreted forecasts. Any averaging aggregator is also constrained to
the convex hull of the individual forecasts, which further contradicts the interpreted signal framework \citep{parunak2013characterizing} and can be far from optimal on many datasets.

\subsection{Empirical Approaches}
\label{ss:empirical}

If one is not concerned with theoretical justification, an obvious
approach is to perturb one of these estimators and observe whether the
adjusted estimator performs better on some data set of interest.  Given that the measurement error
framework produces under-confident aggregators, a popular adjustment is to {\em extremize}, that is, to shift the average aggregates closer to the
nearest extreme (either zero or one).  For instance,~\citet{Ranjan08} 
extremize $\pb$ with the CDF of a
beta distribution; \citet{satopaa} use a logistic
regression model to derive an aggregator that extremizes $\plog$; \citet{baron2014two} give two intuitive justifications for
extremizing and discuss an extremizing technique that has previously
been used by a number of investigators (\citealt{Erev1994,
shlomi2010subjective}; and even \citealt{karmarkar1978subjectively}); \citet{mellers2014psychological} show empirically that extremizing can
improve aggregate forecasts of international events.
%%%\citet{turner2013forecast} and \citet{Ariely00theeffects} 
%%%also mention the benefits of extremizing.

These and many other studies represent the unwieldy position of the
 current state-of-the-art aggregators: they first compute an average 
 based on a model that is
likely to be at odds with the actual process of probability
forecasting, and  then aim to correct the induced bias  via {\em ad hoc}
extremizing techniques.
Not only does this leave something to be desired from an explanatory
point of view, these approaches are also subject to overfitting.
%the problems of
%machine learning, such as overfitting.  
%For example, the amount of 
%extremizing is typically learned by minimizing a
%proper loss function over a separate training set (see \citealt{Gneiting04strictlyproper} for estimation via proper loss functions). This requires repeated realizations
%of a single event and therefore cannot be applied to a single event with an unknown outcome. 
%Furthermore, while 
%extremization works to some degree,
%
Most importantly, these techniques provide little
insight beyond the amount of extremizing itself and hence lack a clear direction of continued improvement.
The present paper aims to remedy this situation by explaining extremization with the aid of a theoretically based estimator, namely the oracular aggregator. 
%
%producing theoretically based estimators which
%explain extremization. 

%The present paper aims to remedy this situation.  
%Under a simple instance of the partial information framework, the optimal amount of extremization can be quantified
%with a single number that follows a Cauchy distribution.  Given that
%this distribution depends on the structure of information among the
%forecasters, it establishes a link between the information structure
%and the amount of extremizing.  This leads to three main results:  First, the modal
%amount of extremizing is increasing in the level of information
%diversity and the total amount of information among forecasters.
%Second, no matter how information is distributed among the forecasters,
%extremizing $\probit$ is more likely to be beneficial
%than harmful (except in the degenerate situation where all forecasts
%are the same). Third, the structure of the information overlap establishes a spectrum of aggregators, ranging from
%average- (most information is public) to voting-like (most information
%is private) techniques.

\section{THE GAUSSIAN PARTIAL INFORMATION MODEL}
\label{sec:model}
\subsection{Motivating Examples}
A central component of the partial information models is the structure of the
information overlap that is assumed to hold among the individual
forecasters.
%Partial information models are sensitive to the structure of the
%information overlap that is assumed to hold among the individual
%forecasters. Model sensitivity is useful if it reacts
%to important inherent features of the problem, but harmful if it adds
%noise that is not reflected in the actual best response to the data.
It therefore behooves us to begin with some simple examples to show
that the optimal aggregate is not well defined without
assumptions on the information structure among the forecasters.

\begin{example}
\label{FirstExample}
%Suppose two forecasters both report a probability of $2/3$ for some
%event.  If they are seeing the same evidence, then the optimal aggregate
%forecast is $2/3$.  If they are seeing different evidence,
%then clearly the combined evidence should give an aggregate forecast
%somewhat greater than $2/3$.  To make the situation more concrete, c
Consider a basket containing a fair coin and a two-headed coin. Two forecasters are asked to predict whether a coin chosen at random is in fact
two-headed. Before making their predictions, the forecasters observe the result of a single flip of
the chosen coin.  Suppose the flip comes out HEADS. Based on this observation, the correct Bayesian probability estimate is  $2/3$.
 If both forecasters see the
result of the same coin flip, the optimal aggregate
 is again $2/3$. On the other hand, if they observe different (conditionally independent) flips of the same coin, the optimal aggregate 
is $4/5$.
\end{example}
In this example, it is not possible to distinguish between the two different information structures simply based on the given predictions,
%. If the forecasters had participated in a long sequence of independent replications of the given task, it may have been possible to eventually ascertain whether they have the same information or not. Similarly, if their forecasts had disagreed, it would have been easy to conclude that they did not observe the same coin flip.  However, in the given situation, there
%was no way to know, 
and neither $2/3$ nor $4/5$ can be said to be a
better choice for the aggregate forecast.  Therefore, we conclude that it is necessary to incorporate an assumption as to the structure of the information
overlap, and that the details must be informed by the particular
instance of the problem. The next example shows that even if the forecasters observe marginally independent events, further details
in the structure of information can still greatly affect the
optimal aggregate forecast.

\begin{example}
Let $\Omega = \{ A,B,C,D \} \times \{ 0,1 \}$ be a probability space
with eight points.  Consider a measure $\mu$ that assigns
probabilities $\mu (A,1) = a/4, \mu (A,0) = (1-a)/4$, $\mu (B,1) =
b/4, \mu (B,0) = (1-b)/4$, and so forth. Define two events 
\begin{align*}
S_1 &= \{(A,0),(A,1),(B,0),(B,1) \},\\
S_2 &= \{(A,0),(A,1),(C,0),(C,1) \}.
\end{align*}
 Therefore, $S_1$ is the event that the first coordinate is
$A$ or $B$, and $S_2$ is the event that the first coordinate
is $A$ or $C$. Consider two forecasters and suppose Forecaster $i$ observes $S_i$. Therefore the $i$th Forecaster's information set is
given by the $\sigma$-field $\F_i$ containing $S_i$ and its
complement. Their $\sigma$-fields are independent. Now, let
$G$ be the event that the second coordinate is~1.  Forecaster~1
reports $p_1 = \P(G | \mathcal{F}_1) = (a+b)/2$ if $S_1$ occurs;
otherwise, $p_1 = (c+d)/2$.  Forecaster~2, on the other hand, reports
$p_2 = \P(G | \mathcal{F}_2) = (a+c)/2$ if $S_2$ occurs; otherwise,
$p_2 = (b+d)/2$.  If $\ee$ is added to $a$ and $d$ but subtracted from
$b$ and $c$, the forecasts $p_1$ and $p_2$ do not change, nor does it
change the fact that each of the four possible pairs of forecasts has
probability $1/4$.  Therefore all observables are invariant under
this perturbation.  If Forecasters $1$ and $2$ report $(a+b)/2$ and
$(a+c)/2$, respectively, then the aggregator knows, by considering the
intersection $S_1 \cap S_2$, that the first coordinate is $A$.
Consequently, the optimal aggregate forecast is $a$, which is most
definitely affected by the perturbation.
\end{example}

This example shows that the aggregation problem can be affected
by the fine structure of information overlap.  It is, however, unlikely
that the structure can ever be known with the precision postulated in
this simple example.  Therefore it is necessary to make reasonable
assumptions that yield plausible yet generic information structures.

\subsection{Gaussian Partial Information Model}
\label{ss:Gaussian}
%This section makes the probability model described in Section \ref{PIFintro} accessible to practice.
The central component of the Gaussian model is a pool of information particles. Each particle, which can be interpreted as representing the smallest unit of information, is either positive or negative. The positive particles provide evidence in favor of the event $A$, 
%For instance, a subset of positive particles may represent a news article with  evidence that $A$ is likely to happen. 
while the negative particles provide evidence against $A$. Therefore, if the overall sum (integral) of the positive particles is larger than that of the negative particles, the event $A$ happens; otherwise, it does not. Each forecaster, however, observes only the sum of some subset of the particles. Based on this sum, the forecaster makes a probability estimate for $A$. This is made concrete in the following model that  represents the pool of information with the unit interval and generates the information particles from a Gaussian process. 

\begin{quote}
{\bf The Gaussian Model.} Identify the pool of information with the unit interval $S = [0,1]$. Consider a centered Gaussian process $\{X_B\}$ that is defined on a probability space $(\Omega
, \F , \P)$ and indexed by the Borel subsets $B \subseteq S$ such that $\Cov (X_B , X_{B'}) = |B \cap B'|$.  Such a process can be constructed, for example, by considering a standard Brownian motion process $Y(t)$ on $[0,1]$, and defining $X_B$ as the variation of $Y$ over $B$.
%In other words, the
%unit interval $S$ is endowed with Gaussian white noise, and $X_B$ is the
%total of the white noise in the Borel subset $B$. 
 Let $A$ denote the
event that the sum of all the information is positive: $A := \{ X_S > 0 \}$.
For each $i = 1, \dots, N$, let $B_i$ be some Borel subset of $S$, and
define the corresponding $\sigma$-field as $\F_i := \sigma (X_{B_i})$. Forecaster $i$ then predicts $p_i :=
\E (\one_A \| \F_i)$.
\end{quote}
%Such a process can be constructed, for instance, by considering a standard Brownian motion $Y(t)$ on $S = [0,1]$, and define $X_{B}$ as the variation of $Y$ over $B \subseteq S$. More specifically, for any interval $I = [a,b]$, let $X_{I} = Y(b) - Y(a)$. Given that $B$ is a Borel set, it can be expressed as the union of (potentially countably many) disjoint intervals. That is, 
The Gaussian model can be motivated by recalling the interpreted signal
model of~\citet{broomell2009experts}. They assume that
Forecaster~$i$ forms an opinion based on $L_i (Z_1 , \ldots , Z_r)$,
where each $L_i$ is a linear function of observable quantities or cues
$Z_1 , \ldots , Z_r$ that determine the outcome of $A$.  
%Proposing a linear model for subjective
%interpretation seems quite natural and is in fact the only such
%postulate we know to have been suggested for interpreted signals.  
If
the observables (or any linear combination of them) are independent
and have small tails, then as $r \to \infty$, the joint distribution
of the linear combinations $L_1 , \ldots , L_N$ will be asymptotically
Gaussian. 
Therefore, given that the number of cues in a real-world setup is likely to be large, it makes sense
to model the forecasters' observations as jointly Gaussian. The remaining component, namely the covariance structure of the joint distribution is then motivated by the partial information framework. Of course, other distributions, such as the $t$-distribution, could be considered. However, given that both the multivariate and conditional  Gaussian distributions have simple forms, the Gaussian model offers potentially the cleanest entry into the issues at hand. 
 
% Other partial information models have been considered in the past. 
 
Overall, modeling the forecasters' predictions with a Gaussian distribution is rather common. For instance, \cite{Bacco} consider a model of two forecasters whose estimated log-odds follow a joint Gaussian distribution. The predictions are assumed to be based on different information sets; hence, the model can be viewed as a partial information model. Unfortunately, as a specialization of the partial information
framework, this model is a fairly narrow due to its detailed
assumptions and extensive computations. The end result is a rather restricted aggregator of two probability forecasts. On the contrary, the Gaussian model sustains flexibility by specializing the framework only as much as is necessary. The 
following enumeration provides further interpretation and clarifies which
aspects of the model are essential and which have little or
no impact.

%\marginpar{Read the paper by Mario Di Bacco}

\begin{enumerate}[(i)]
\item {\bf Interpretations.} It is not necessary to assume anything 
about the source of the information.  For instance, the information 
could stem from survey research, records, books, 
interviews, or personal recollections.  All these details have 
been abstracted away. \vspace{-0.5em}

\item {\bf Information Sets.} The set $B_i$ holds the information used by
 Forecaster $i$, and the covariance $\Cov (X_{B_i} , X_{B_j}) = |B_{i} \cap B_{j}|$
represents the information overlap between Forecasters $i$ and
$j$.
%\item In the interpreted signal framework, the set $B_i$ represents 
%the linear regressor used by Forecaster~$i$, and the covariance 
%structure represents degrees of similarities between regressors 
%of different forecasters.  
Consequently, the complement of $B_i$ holds information not used by
Forecaster~$i$.  
No assumption is necessary as to whether this
information was unknown to Forecaster~$i$ instead of known but not
used in the forecast. \vspace{-0.5em}

\item {\bf Pool of Information.} First, the pool of
information potentially available to the forecasters is the white
noise on $S = [0,1]$. The role of the unit interval
is for the convenient specification of the sets $B_i$.
The exact choice is not relevant, and
any other set could have been used. The unit interval, however, is a
natural starting point that 
%provides 
%an alternative interpretation of $|B_j|$ as marginal probabilities for some $N$ events, $|B_i \cap B_j|$ as their pairwise joint probabilities, $|B_i \cap B_j \cap B_k|$ as their three-way joint probabilities,
%and so forth.  This interpretation is particularly useful in analysis as it
%
links the
information structure to many known results in combinatorics and geometry (see, e.g.,  Proposition \ref{CorrelationPolytope}). Second, there is
no sense of time or ranking of information within the
pool. Instead, the pool is a collection of information, where each
piece of information has an {\em a priori} equal chance to contribute
to the final outcome.  Quantitatively, information is parametrized
by the length measure on $S$. \label{item:pool} \vspace{-0.5em}

\item {\bf Invariant Transformations.}  From the empirical point of
view, the exact identities of the individual sets $B_i$ are
irrelevant.  All that matters are the covariances $\Cov \left(X_{B_i}
, X_{B_j}\right) = |B_i \cap B_j|$.  The explicit sets $B_i$ are only useful in the analysis, e.g., when computing the oracular aggregator.

\item {\bf Scale Invariance.} The model is invariant under rescaling,
replacing $S$ by $[0,\lambda]$ and $B_i$ by $\lambda B_i$.  Therefore,
the actual scale of the model (e.g., the fact that the covariances of
the variables $X_B$ are bounded by one) is not relevant. \vspace{-0.5em}

\item {\bf Specific vs. General Model.} A specific model requires a
choice of an event $A$ and Borel sets $B_i$.
This might be done in several ways: a) by choosing them in advance,
according to some criterion; b) estimating the parameters $\P(A)$,
$|B_i|$, and $|B_i \cap B_j|$ from data; or c) using a Bayesian model
with a prior distribution on the unknown parameters.  This paper
focuses mostly on a) and b) but discusses c) briefly in
Section~\ref{discussion}.  Section \ref{extremizing} provides one
result, namely Proposition~\ref{positiveProbThm} that holds for any (nonrandom) choices of the sets $B_i$. \vspace{-0.5em}
\label{item:specific}

\item {\bf Choice of Target Event.} There is one substantive
assumption in this model, namely the choice of the half-space for the event $A$. Choosing  $\{
X_S > t \}$ for some $t \in \mathbb{R}$ makes the prior
probability equal to $\P(A) = 1-\Phi(t)$.  The current paper defers the analysis of $t \neq 0$ to future work and focuses on the
centered model for simplicity. 
%but also because 
%There is one substantive
%assumption in this model, namely the choice of the half-space $\{ X_S
%> 0 \}$ for the event $A$.   Changing this event results
%in a non-isomorphic model. 
%This would make the prior
%probability equal to $\P(A) = 1-\Phi(t) \neq 1/2$. 
%The current choice 
Furthermore, choosing $t = 0$ implies a prior probability $\P(A) = 1/2$, which seems as
uninformative as possible and therefore provides a natural starting
point. Note that specifying a prior distribution for $A$ cannot be avoided as
long as the model depends on a probability space. This includes essentially any probability
model for forecast aggregation. \vspace{-0.5em}
\label{item:choice}

%
%
%\item {\bf Centering.} One could choose a non-central half-space $\{
%X_S > t \}$ with $t \neq 0$ for the event $A$.  This would make the prior
%probability equal to $\P(A) = 1-\Phi(t) \neq 1/2$. The current paper, however, focuses on the
%centered model for simplicity but also for the following reason.  
%%The
%Gaussian assumption is most often true in practice when the
%observations are centered.  For example, if the electorate
%is broken into a dozen demographic segments, the portion of voters in
%each segment that vote for a given candidate does not follow a
%Gaussian distribution, but the difference between this portion and the
%historically expected or predicted portion may.  A centered partial
%information model is then quite realistic when forecasting the event
%of a positive deviation. 
% The political futures website, Intrade,
%during its operation, was in fact well stocked with centered events. 
%%For instance, the website might have inquired whether 
%% the number of representatives elected from a
%%given party will exceed a threshold that had been adjusted to
%%bring the prior probability near $1/2$.  
%Sports betting websites also operate
%almost exclusively in this mode.
\label{item:centered}
\end{enumerate}

\subsection{Preliminary Observations}
\label{prelim}
%The remaining details of the Gaussian model are best described by first considering only two Forecasters $1$ and $2$ . Suppose they observe respective $\delta_1$ and
%$\delta_2$ portions of the Gaussian process and that their information overlap has size $\rho$. That is, $|B_1| = \delta_1$, $|B_2| = \delta_2$, and $|B_1 \cap B_2| = \rho$.
%\begin{figure}[t]
%%   \hspace{-2em}
%   \includegraphics[width = \textwidth]{N=2} % requires the graphicx package
%   \caption{Illustration of the Gaussian Partial Information Model with $2$ Forecasters.}
%   \label{diagram2}
%\end{figure}
%Figure \ref{diagram2} illustrates this setup.  In this diagram, the
%Gaussian process has been partitioned into four parts based on the
%forecasters' information sets: $U = X_{B_1 / B_2}$, $M = X_{B_1 \cap B_2}$, $V = X_{B_2 / B_1}$, and $W = X_{(B_1 \cup B_2)^c}$. The partition illustrates the additive nature of the information pool. In particular, $X_{B_1} = U + M$,  $X_{B_2} = M + V$, $X_S = U+M+V+W$, where $U, V, M, W$ are independent Gaussian random variables with respective variances
%$\delta_1-\rho$, $\delta_2-\rho$, $\rho$, and $1+\rho-\delta_1 -
%\delta_2$. The joint distribution of $X_{S}$, $X_{B_1}$, and $X_{B_2}$ is a
%multivariate Gaussian distribution:
%\begin{align}
%\left(\begin{matrix} X_S \\ X_{B_1}\\ X_{B_2} \end{matrix}\right) 
% &\sim \mathcal{N}\left(
% \boldsymbol{0},  \left(\begin{matrix} 
%1 & \delta_1 & \delta_2\\
%\delta_1 & \delta_1 &\rho\\
%\delta_2 & \rho & \delta_2
% \end{matrix}\right)\right) \label{twoforecasters}
%\end{align}
 The Gaussian process exhibits additive behavior that aligns well with the intuition of an information pool. To see this, consider a finite partition of the full information $\{ C_\v := \cap_{i \in \v} B_i \setminus \cup_{i \notin \v} B_i : \v \subseteq \{1, \dots, N\}\}$. Each subset $C_\v$ represents a set of information particles such that $B_i = \bigcup_{\v \ni i} C_\v$ and $X_{B_i} = \sum_{\v \ni i} X_{C_\v}$. Therefore $X_{B}$ can be regarded as the sum of the particles in the subset $B \subseteq S$, and different $X_{B}$'s relate to each other in a manner that is consistent with this interpretation. 
\begin{figure}[t]
\centering
\begin{minipage}[t]{0.49\textwidth}
  \centering
  \raisebox{0.073\height}{\includegraphics[width=\linewidth]{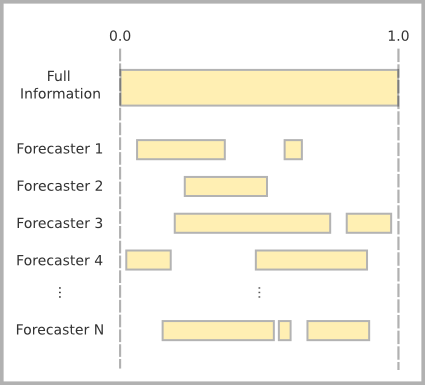}}
  \captionof{figure}{Illustration of Information Distribution among $N$ Forecasters. The bars leveled horizontally with Forecaster $i$ represent the information set $B_i$.}
%  Each forecasters can observe any subset of the full information $[0,1]$.}
  \label{diagramN}
\end{minipage}%
\hspace{0.5em}
\begin{minipage}[t]{0.49\textwidth}
  \centering
  \includegraphics[width=0.973\linewidth]{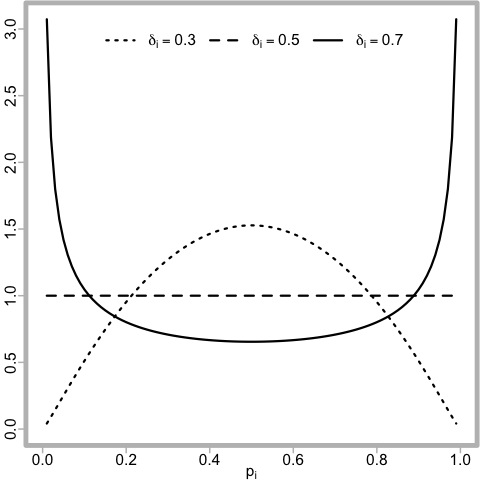}
  \captionof{figure}{Marginal Distribution of $p_i$ under Different Levels of 
$\delta_i$. The more the forecaster knows, 
the more the forecasts are concentrated around the extreme 
points zero and one.}
  \label{marginals}
\end{minipage}
\end{figure}
The relations among the relevant variables are summarized by a multivariate Gaussian distribution:
\begin{align}
\left(\begin{matrix} X_S \\ X_{B_1}\\ \vdots \\ X_{B_N} \end{matrix}\right) &\sim \mathcal{N}\left( 
%\boldsymbol{X} &\sim \mathcal{N}\left( 
%\left(\begin{matrix} 
%\mu_1 \\ \boldsymbol{\mu}_2
% \end{matrix}\right) =
 \boldsymbol{0}, \left(\begin{matrix} 
\Sigma_{11} & {\bf \Sigma}_{12}\\
{\bf\Sigma}_{21} & {\bf \Sigma}_{22}\\
 \end{matrix}\right) 
 =
 \left(\begin{array}{c | c c cc }
1 & \delta_1 & \delta_2 & \dots & \delta_N  \\ \hline
\delta_1 & \delta_1 &\rho_{1,2} & \dots & \rho_{1,N}   \\ 
\delta_2 & \rho_{2,1} & \delta_2 & \dots & \rho_{2,N}  \\ 
\vdots & \vdots & \vdots & \ddots & \vdots  \\ 
\delta_N & \rho_{N,1} & \rho_{N,2} & \dots & \delta_N\\ 
 \end{array}\right)\right),  \label{Nforecasters}
\end{align}
where $|B_i| = \delta_i$ is the amount of information used by Forecaster $i$, and $|B_i \cap B_j| = \rho_{ij} = \rho_{ji}$ is the amount of information overlap between Forecasters $i$ and $j$. One possible instance of this setup is illustrated in Figure \ref{diagramN}. Note that $B_i$ does not have to be a contiguous subset of $S$.  Instead, each forecaster can use any Borel measurable
subset of the full information. 

Under the Gaussian model, the sub-matrix ${\bf \Sigma}_{22}$ is sufficient for the information structure. Therefore the exact identities of the Borel sets do not matter, and learning about the
information among the forecasters is equivalent to estimating a
covariance matrix under several restrictions.  In particular, if the
information in ${\bf\Sigma}_{22}$ can be translated into a diagram
such as Figure \ref{diagramN},
the matrix ${\bf\Sigma}_{22}$ is called \textit{coherent}.  
This property is made precise in the following proposition. 
%The following proposition describes the space of coherent information structures.
 The proof of this and other propositions are deferred to Appendix A of the Supplementary Material.

\begin{proposition}
\label{CorrelationPolytope}
The overlap structure ${\bf\Sigma}_{22}$ is coherent if and only
if 
%\begin{align*}
${\bf\Sigma}_{22} \in \COR(N) := \conv\left\{
\boldsymbol{x}\boldsymbol{x}' : \boldsymbol{x} \in
\{0,1\}^N\right\}$,
%&= \left\{ \sum_{i=1}^{2^N} \lambda_i  \right\}\\
%\end{align*}
where $\conv\{\cdot\}$ denotes the convex hull and $\COR(N)$ is known as the correlation
polytope. It is described by $2^N$
vertices in dimension $\text{dim}(\COR(N)) = \binom{N+1}{2}$.
\end{proposition}
The correlation polytope has a very complex description in terms of half-spaces. In fact, complete descriptions of the facets of $\COR(N)$ are only known for $N \leq 7$ and conjectured for  $\COR(8)$ and $\COR(9)$ 
%\citep{ziegler2000lectures, christofsmapo, pitowsky1991correlation}
\citep{ziegler2000lectures}. 
%For instance, 
%\begin{align*}
%%\COR(2) &= 
%%\left\{ {\bf\Sigma}_{22} : 
%%\begin{array}{ll}
%% 0 \leq \rho_{12} \leq \min(\delta_1, \delta_2)\\
%% \delta_1 + \delta_2 - \rho_{12} \leq 1
%%\end{array}
%%\right\}\\
%\COR(3) &= 
%\left\{ {\bf\Sigma}_{22} : 
%\begin{array}{ll}
% 0 \leq \rho_{ij} \leq \min(\delta_i, \delta_j)\\
% \delta_i + \delta_j - \rho_{ij} \leq 1\\
% \delta_i - \rho_{ij} - \rho_{ik} + \rho_{jk} \geq 0\\
% \delta_1 + \delta_2 + \delta_3 - \rho_{12} - \rho_{13} - \rho_{23} \leq 1
%\end{array}
%\right\},
%\end{align*}
%%\begin{cases}
%% 0 \leq \rho_{12} \leq \min(\delta_1, \delta_2)\\
%% \delta_1 + \delta_2 - \rho_{12} \leq 1
%%\end{cases}
%%\end{align*}
%%and $\COR(3)$ with 
%%\begin{align*}
%%\hspace{5.5em}\begin{cases}
%% 0 \leq \rho_{ij} \leq \min(\delta_i, \delta_j)\\
%% \delta_i + \delta_j - \rho_{ij} \leq 1\\
%% \delta_i - \rho_{ij} - \rho_{ik} + \rho_{jk} \geq 0\\
%% \delta_1 + \delta_2 + \delta_3 - \rho_{12} - \rho_{13} - \rho_{23} \leq 1
%%\end{cases}
%%\end{align*}
%%for $i,j,k \in \{1,2,3\}$ such that $i \neq j$, $i \neq k$, and $j \neq k$.  
%$\COR(5)$ has 56 known facets, and $\COR(9)$ has at least 12,246,651,158,320 facets. Because the unconstrained quadratic
%0-1 problem is NP-hard, it is probably hopeless to find a complete
%list of linear inequalities to describe $\COR(N)$ 
%\citep{deza1997geometry}.
 Fortunately, previous literature has introduced both linear and semidefinite relaxations of $\COR(N)$ \citep{laurent1997connections}. Such relaxations together with modern optimization techniques and sufficient data can be used to estimate the information structure very efficiently. This, however, is not in the scope of this paper and is therefore left for subsequent work. 

%To link the information pool with the forecasts, observe that $X_S | X_{B_i} \sim \mathcal{N}\left(X_{B_i}, 1-\delta_j\right)$. 

The multivariate Gaussian distribution (\ref{Nforecasters}) relates to the forecasts by
\begin{align}
p_i &= \P\left(A | \mathcal{F}_{i}\right) = \P\left(X_S > 0 | X_{B_i}\right) = \Phi\left( \frac{X_{B_i}}{\sqrt{1-\delta_i}}\right). \label{indFore}
\end{align}
%This forecast is a conditional expectation of $\one_A$ given $\mathcal{F}_i$ and hence  calibrated. Its
%The marginal distribution of $p_i$ can be computed by first denoting the probit score with $P_{i} := \Phi^{-1}(p_i) = X_{B_i}/\sqrt{1-\delta_i}$. The Jacobian for the map $P_{i} \to \Phi(P_i)$ is $J(P_i) = (2\pi)^{-1/2} \exp \left( - P_i^2/2   \right)$. If $h(P_i)$ denotes the Gaussian density of $P_i \sim \mathcal{N}\left(0, \delta_i / (1-\delta_i)\right)$, 
The marginal density of $p_i$,
\begin{align*}
 m\left(p_i | \delta_i \right)
%  &= h(P_i) J(P_i)^{-1} \big|_{P_i = \Phi^{-1}(p_i)} 
 = \sqrt{\frac{1-\delta_i}{\delta_i}} \exp 
   \left\{ \Phi^{-1}(p_i)^2 \left(1-\frac{1}{2 \delta_i} \right) \right\}, 
\end{align*}
 has very intuitive behavior: it is uniform on $[0,1]$ if $\delta_i = 1/2$, but becomes 
unimodal with a minimum (maximum) at $p_i = 1/2$ when $\delta_i > 1/2$ ($\delta_i < 1/2$).  As $\delta_i \to 0$, $p_i$ converges to a point mass at $1/2$. On the other hand, as $\delta_i \to 1$, $p_i$ converges to
a correct forecast whose distribution has atoms of weight $1/2$ at
zero and one. Therefore a forecaster with no information ``withdraws'' from the problem by predicting a non-informative probability $1/2$ while a forecaster with full information always predicts the correct outcome with absolute certainty. Figure \ref{marginals} illustrates the marginal
distribution when $\delta_i$ is equal to $0.3$, $0.5$, and $0.7$.

%\begin{figure}[t]
%\centering
%	\hspace{0em}\includegraphics{LegendMarginal}
%
% \includegraphics[width= 0.55\textwidth]{Marginals}
%   \caption{The Marginal Distribution of $p_i$ under Different Levels of 
%$\delta_i$.  The more the forecaster knows, i.e., the higher $\delta_i$ is, 
%the more the probability forecasts are concentrated around the extreme 
%points 0 and~1.}
%\label{marginals}
%\end{figure}

\section{PROBABILITY EXTREMIZING}
\label{extremizing}
\subsection{Oracular Aggregator for the Gaussian Model}
\label{oracular}
%We show that, under two specific Gaussian partial 
%information models, the oracle, on average,
%extremizes the probit aggregator. 
Recall from Section \ref{PIFintro} that the oracular aggregator is the
conditional expectation of $\one_A$ given all the forecasters' information. Under the Gaussian model, this can be
emulated with a hypothetical oracle forecaster whose information set is
$B' := \bigcup_{i=1}^N B_i$.  
%Appending this to the multivariate
%Gaussian distribution~(\ref{Nforecasters}) gives
%\begin{align}
%\left(\begin{matrix} X_S \\ X_{B'} \\ X_{B_1}\\ \vdots \\ X_{B_N} 
% \end{matrix}\right) &\sim \mathcal{N}\left( 
% \boldsymbol{0}, 
%% \left(\begin{matrix} 
%%\Sigma_{11}' & \Sigma_{12}'\\
%%\Sigma_{21}' & \Sigma_{22}\\
%% \end{matrix}\right) 
%% =
%% 
% \left(\begin{array}{c c| c c cc }
%1 & \delta' & \delta_1 & \delta_2 & \dots & \delta_N  \\ 
%\delta' & \delta' & \delta_1 & \delta_2 & \dots & \delta_N  \\ \hline
%\delta_1& \delta_1 & \delta_1 &\rho_{1,2} & \dots & \rho_{1,N}   \\ 
%\delta_2 & \delta_2 &\rho_{2,1} & \delta_2 & \dots & \rho_{2,N}  \\ 
%\vdots &\vdots & \vdots & \vdots & \ddots & \vdots  \\ 
%\delta_N &\delta_N & \rho_{N,1} & \rho_{N,2} & \dots & \delta_N\\ 
% \end{array}\right)\right), \label{oracleN}
%\end{align}
%where $X_{B'}$ is the information known to the oracle and $\delta' =
%|B'|$.
 The oracular aggregator is then nothing more than the probability forecast made by the oracle. That is,
 \begin{align*}
p' &= \P(A |  \mathcal{F}') = \P(X_S > 0 |  X_{B'}) = \Phi\left( \frac{X_{B'}}{\sqrt{1-\delta'}} \right),
\end{align*}
where $\delta' = |B'|$. Given that the oracle's information set $B'$ cannot be used to reconstruct the individual sets $\{B_i\}_{i=1}^N$, some potentially relevant information may appear to have been lost. Under the Gaussian model, however, only the total variation over $B'$ is relevant to aggregation. The next proposition shows that $X_{B'}$ contains all the information in $\{X_{B_i}\}_{i=1}^N$ and hence leads to an actual oracular aggregator. 
\begin{proposition} \label{condInd}
The event $A$ is conditionally independent of the collection $\{X_{B_i}\}_{i=1}^N$ given $X_{B'}$
\end{proposition}
%This construction relies on the fact that $A$ is conditionally independent of the collection $\{X_{B_i}\}_{i=1}^N$ given $X_{B'}$.
%; that is, the sets $\{B_i\}_{i=1}^N$ hold no information that is not in $B'$. 

The oracular aggregator provides a reference point that allows us to identify information structures under which other aggregation techniques perform relatively well. In particular, if an aggregator is likely to be near $p'$ under a given ${\bf\Sigma}_{22}$, then that information structure reflects favorable conditions for the aggregator. 
This ideas is used in the following subsections to develop intuition about probability extremizing. 

\subsection{General Information Structure}
%This section begins the analysis of probability extremizing by making some preliminary observations under the general information structure. First, however, recall that 
%Even though the results are mainly targeted at forecasting practitioners, the discussion serves as an illustration on how the oracular aggregator can be used to benchmark and understand other aggregation techniques. 

A probability $p$ is said to be \textit{extremized} by another probability $q$ if and only
if $q$ is closer to zero when $p \leq 1/2$ and closer to one when $p
\geq 1/2$. This translates to the probit scores as follows: $q$ extremizes $p$ if and only if  $\Phi^{-1}(q)$ is on the same side but further away from zero than $\Phi^{-1}(p)$. The amount of (multiplicative) extremization can then be quantified with the {\em probit extremization ratio} defined as 
 $\alpha(q,p) := \Phi^{-1}(q) / \Phi^{-1} (p)$.

 Given that no aggregator can improve upon the oracular aggregator, it provides an ideal
reference point for analyzing extremization. This section specifically uses it to study
extremizing of $\probit$ because a) it is arguably more
reasonable than the simple average $\bar{p}$; and b) it is very
similar to $\plog$ but results in cleaner
analytic expressions. Therefore, of particular interest is the special case $\alpha(p', \probit)  = P' \big/ \left(\frac{1}{N}\sum_{i=1}^N P_{i} \right)$, where $P' = \Phi^{-1}(p')$. From now on, unless otherwise stated, this expression is referred simply with
$\alpha$. 
Therefore, the probit opinion pool $\probit$
requires extremization if and only if $\alpha > 1$, and the larger $\alpha$ is, the more $\probit$ should be extremized. 

Note that $\alpha$ is a random quantity that spans the entire real line;
that is, it is possible to find a set of forecasts and an information structure for any possible value of $\alpha \in
\mathbb{R}$.  Evidently, extremizing is not guaranteed to always
improve $\probit$.  To understand when extremizing is
likely to be beneficial, the following proposition provides the probability
distribution of $\alpha$.

\begin{proposition}
\label{positiveProbThm}
The law of the extremization ratio $\alpha$ is a Cauchy with
parameters $x_0$ and $\gamma$, where the location parameter $x_0$ is
at least one, equality occurring only when $\delta_i = \delta_j$ for all
$i \neq j$. Consequently, if $\delta_i \neq \delta_j$ for some
$i \neq j$, then the probability that $\probit$
requires extremizing $\P\left(\alpha > 1 | {\bf\Sigma}_{22}, \delta'\right)$
is strictly greater than $1/2$.
%\textcolor{red}{Must add the Cauchy distribution parameter somewhere}
\end{proposition}
\noindent
This proposition shows that, on any non-trivial problem, a small
perturbation in the direction of extremizing is more likely to improve
$\probit$ than to degrade it.  This partially explains
why extremizing aggregators perform well on large sets of real-world
prediction problems.  It may be unsurprising after the fact, but the
forecasting literature is still full of articles that perform
probability averaging without extremizing. The next two subsections examine special cases in which more detailed computations can be performed.

%If qualitative analysis is sufficient, it may be possible to avoid bound approximations by focusing on how $\delta'$ changes as a function of other parameters instead. For instance, $\delta'$ increases in $N$ and $\delta_{i}$ but decreases in $\rho_{ij}$. These partial effects are utilized in Section \ref{compound}. The next subsection, on the other hand, examines two special cases in which the oracular aggregator coincides with the revealed aggregator. The main goal is to shed light on extremization in
%real-world forecasting setups.

\subsection{Zero and Complete Information Overlap}
\label{disjoint}
%This section examines two extreme cases: a) the forecasters share
%no information; and b) the forecasters share all their
%information. First, consider the latter case where all the information
%sets are the same,
If the forecasters use the same information, i.e.,   $B_{i} = B_j$ for all $i \neq j$, their forecasts are identical, $p' = p'' = \probit$, 
%all
%the probability forecasts $\{p_i : i = 1, \dots, N\}$ are the same,
%and the probit opinion pool, the oracular aggregator, and the revealed
%aggregator equal to this common probability forecast. 
and no
extremization is needed. Therefore,
given that the oracular aggregator varies
smoothly over the space of information structures, 
averaging
techniques, such as $\probit$, can be expected to work
well when the forecasts are based on very similar sources of
information. This result is supported by the fact that the
measurement error framework, which essentially describes the
forecasters as making numerous small mistakes while applying the same
procedure to the same data (see Section \ref{ss:measurement}),
results in averaging-based aggregators.

If, on the other hand, the forecasters have zero information overlap, i.e.,  $|B_{i} \cap B_{j}| = 0$ for all $i \neq j$, the information structure ${\bf\Sigma}_{22}$ is
diagonal 
%Such a form is likely to arise if a team of forecasters
%strategically decide to access and study disjoint sources of
%information.
and 
%Given that under non-overlapping information 
%$\delta' = \sum_{j=1}^N \delta_j$ and $X_{B'} = \sum_{j=1}^N X_{B_j}$, 
 \begin{align*}
p' &= p'' =  \Phi\left( \frac{\sum_{i=1}^N X_{B_i}}
  {\sqrt{1- \sum_{i=1}^N \delta_i}} \right),
\end{align*}
where the identities $\delta' = \sum_{i=1}^N \delta_i$ and $X_{B'} = \sum_{i=1}^N X_{B_i}$ result from the additive nature of the Gaussian process (see Section \ref{prelim}).  This aggregator can be described in two steps: First, the numerator conducts voting, or range voting to be more specific, where the votes are weighted according to
the importance of the forecasters' private information. Second, the
denominator extremizes the consensus according to the total
amount of information in the group. This clearly leads to very extreme forecasts. Therefore
more extreme techniques can be expected to work well
when the forecasters use widely different information sets.

%Combining this with our earlier discussion leads to the following observation.
The analysis suggests a spectrum of aggregators indexed by
the information overlap: the optimal aggregator undergoes a smooth
transformation from averaging (low extremization) to voting (high extremization) as
the information overlap decreases from complete to zero overlap.
%This is summarized in the following observation.
%%\begin{observation}
%The information structures index a spectrum of aggregators. 
%That is,
%\begin{center}
%\vspace{-1em}
%\singlespacing
%\begin{tabular}{lcl}
%%\stackrel{\text{ \normalsize full information overlap}}{\stackrel{\text{\normalsize no extremization}}{averaging}} && \Leftrightarrow  && \stackrel{\text{ no information overlap}}{\text{full extremization}}
%%high information overlap & \multirow{2}{*}{$\Leftrightarrow$} & low information overlap\\
%high information overlap & & low information overlap\\
%low extremization & {\Large $\Longleftrightarrow$} & high extremization \\
%averaging  & & voting\\
%\end{tabular}
%\end{center}
%%\end{observation}
This observation gives qualitative guidance in real-world settings
where the general level of overlap can be said to be high or low.  For
instance, predictions from forecasters working in
close collaboration can be averaged while predictions from forecasters strategically accessing and studying disjoint sources of
information should be
aggregated via more extreme techniques such as voting. See \citealt{parunak2013characterizing} for a discussion of voting-like techniques. For a concrete illustration, recall  Example \ref{FirstExample} where the optimal aggregate changes from $2/3$ (high information overlap) to $4/5$ (low information overlap).

\subsection{Partial Information Overlap}
\label{compound}
%In this section, however, only a lower bound is of interest because it facilitates the analysis of the modal amount of extremization under general $N$.

%where
%\begin{align*}
%\delta'_{min} &= \frac{N \sum_{i=1}^N \delta_i - 2 \sum_{1 \leq i \leq j \leq N} \rho_{ij}}{\lfloor \frac{N+1}{2} \rfloor \lceil \frac{N+1}{2} \rceil }\\
%\delta'_{max} &=  \leq \delta' \leq \frac{N \sum_{i=1}^N \delta_i - 2 \sum_{1 \leq i \leq j \leq N} \rho_{ij}}{N}
%\end{align*}
%Note that both of these bounds reduce to $\delta_1 + \delta_2 -
%\rho_{12}$ when $N=2$. In this section, the lower bound is of particular interest because
%it allows the modal amount of extremization to be bounded from below under general $N$. See \citealt{deza1997geometry} for the further discussion of these and other
%bounds.
%
%\textcolor{red}{It is not really to make this visualizable but to bring this on a higher level. }

To analyze the intermediate scenarios with partial information overlap among the forecasters, it is helpful to reduce the number of parameters in ${\bf\Sigma}_{22}$. A natural approach is to assume compound symmetry, where the
information sets have the same size and that the amount of pairwise overlap
is constant. More specifically, let $|B_{i}| =
\delta$ and $|B_{i} \cap B_{j}| = \lambda \delta$, where $\delta$ is the amount of information used by each forecaster and
$\lambda$ is the overlapping proportion of this information. The resulting information structure is ${\bf\Sigma}_{22} = {\bf I}_N(\delta - \lambda \delta) + {\bf J}_N \lambda \delta$, where ${\bf I}_N$ is the identity matrix and ${\bf J}_N$ is $N \times N$ matrix of ones.
It is
coherent if and only if 
\begin{align}
\delta \in [0,1] \hspace{2em} \text{and} \hspace{2em} \lambda | \delta \in \left[  
   \max \left\{ \frac{N-\delta^{-1}}{N-1}, 0\right\}, 1 \right]. \label{feasible}
    \end{align}
See Appendix A of the Supplementary Material for the derivation of these constraints. 
%\end{align}

%where the open upper bound on $\lambda$ ensures a non-singular ${\bf\Sigma}_{22}$.
Under these assumptions, the location parameter of the Cauchy distribution of $\alpha$
simplifies to
%\begin{align*}
%\alpha &= \frac{X_{B'}}{\frac{1}{N}\sum_{j=1}^N X_{B_j}} \sqrt{\frac{1-\delta}{1-\delta'}} \sim \text{Cauchy}(x_0, \gamma)
%\end{align*}
% with
%\begin{align*}
$x_0 = N/(1+(N-1)\lambda)  \sqrt{(1-\delta)/(1-\delta')}$. 
% &&& \text{ and } &&& \gamma &=  \sqrt{\frac{N(\delta' + \delta' \lambda (N-1) - \delta N)}{\delta (\lambda (N-1) + 1)^2}}\sqrt{\frac{1-\delta}{1-\delta'}}
%\end{align*}
%The location parameter $x_0$ can be lower bounded by replacing
%$\delta'$ with $\delta'_{min}$ when $N > 2$. Making this substitution and recalling
%Proposition \ref{positiveProbThm} gives us
%\begin{align}
%x_{min} := \max\left\{ \frac{N}{1+(N-1)\lambda}  \sqrt{\frac{1-\delta}{1-\delta'_{min}}}, 1 \right\} \leq x_0 \label{bound}
%\end{align}
\begin{figure}[t]
%\centering
%%\hspace*{2em}  $\log(\alpha)$
%\hspace*{1.2em} 	\includegraphics[width=0.973\textwidth, height = 3em]{colorkey} % requires the graphicx package
%\hspace{-1.2em}
        \centering
        \begin{subfigure}[b]{0.5\textwidth}
                \includegraphics[width=\textwidth, height = \textwidth]{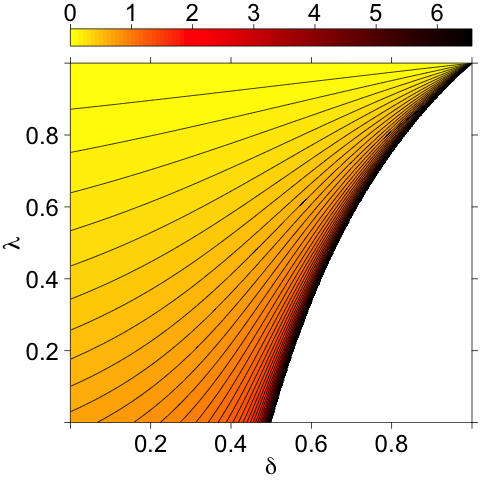}
\caption{$\log(x_0)$}	
\label{xOracle}
        \end{subfigure}%
%\hspace{0.6em}
%        \begin{subfigure}[b]{0.33\textwidth}
%                \includegraphics[width= 0.95\textwidth, height = \textwidth]{ExtremeGamma}
%\caption{$\gamma$}
%\label{gammaOracle}
%        \end{subfigure}
%\hspace{1.3em}
        \begin{subfigure}[b]{0.5\textwidth}
                \includegraphics[width=\textwidth, height = \textwidth]{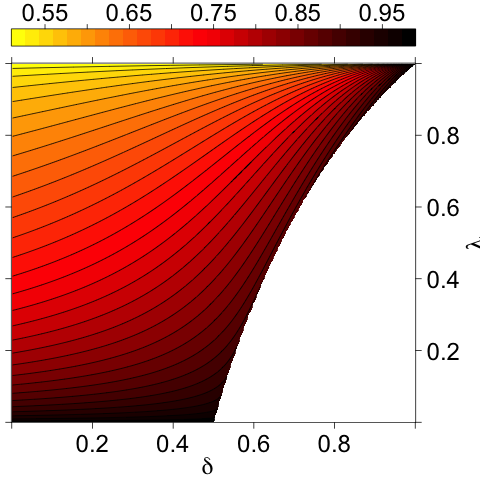}
\caption{$\P(\alpha > 1 | {\bf\Sigma}_{22})$}
\label{probOracle}
        \end{subfigure}

        \caption{Extremization Ratio under Symmetric Information. The amount of extremizing $\alpha$ follows a Cauchy$(x_0, \gamma)$, where $x_0$ is a location parameter and $\gamma$ is a scale parameter. This figure considers  $N = 2$ because in this case $\delta'$ is uniquely determined by ${\bf\Sigma}_{22}$.}
%         experts and shows $\log(x_0)$ and $\P(\alpha > 1 | {\bf\Sigma}_{22})$ under all plausible combinations of $\delta$ and $\lambda$. }
%%        No approximation is needed as $\delta'$ is uniquely determined by ${\bf\Sigma}_{22}$ when $N = 2$.}
        \label{LevelplotsOracle}
\end{figure}
Of particular interest is to understand how this changes as a function of the model parameters. The analysis is somewhat hindered by the unknown details of the dependence between $\delta'$ and the other parameters $N$, $\delta$, and $\lambda$. However, given that $\delta'$ is defined as $\delta' = |\cup_{i=1}^N B_i|$, its value  increases in $N$ and $\delta$ but decreases in $\lambda$. In particular, as $\delta \to 1$, the value of $\delta'$ converges to one at least as fast as $\delta$ because  $\delta' \geq \delta$. Therefore the term $\sqrt{(1-\delta)/(1-\delta')}$ and, consequently, $x_0$ increase in $\delta$. Similarly, $x_0$ can be shown to increase in $N$ but to decrease in $\lambda$. Therefore $x_0$ and $\delta'$ move together, and the amount of extremizing
can be expected to increase in $\delta'$.  As the Cauchy distribution is symmetric around $x_0$, the probability $\P(\alpha > 1 | {\bf\Sigma}_{22})$ behaves similarly to $x_0$ and also increases in $\delta'$.  
%The rate of increase, however, is not the same because it also depends on $\gamma$.  
Figure \ref{LevelplotsOracle} illustrates these relations by plotting both $\log(x_0)$ and $\P(\alpha > 1 | {\bf\Sigma}_{22})$ for $N = 2$ forecasters under all plausible combinations of $\delta$ and $\lambda$. The white space collects all pairs $(\delta,\lambda)$ that do not satisfy (\ref{feasible}) and hence represent incoherent information structures. Note that the results are completely general for the two-forecaster case, apart from the assumption $\delta_1 = \delta_2$. Relaxing this assumption does not change the qualitative nature of the results.
%, namely that extremizing tends to increase in $\delta$ but decrease in $\lambda$.
% Assuming symmetric information among $N = 10$ forecasters, however, is more constraining. In addition, Figure \ref{xOracle10} describes the lower bound $x_{min}$ instead of the exact location $x_0$. Despite these limitations, it illustrates that extremizing can be expected to increase in $N$. 

The total amount of information used by the forecasters $\delta'$, however, does not provide a full
explanation of extremizing.  Information diversity is an
important yet separate determinant.  To see this, observe that fixing $\delta'$ to
some constant defines a curve $\lambda = 2 - \delta'/\delta$ on the two plots in  Figure \ref{LevelplotsOracle}. For
instance, letting $\delta' = 1$ gives the boundary curve on the right
side of each plot.  This curve then shifts inwards and rotates slightly
counterclockwise as $\delta'$ decreases.  At the top end of each curve
all forecasters use the total information, i.e., $\delta =
\delta'$ and $\lambda = 1.0$.  At the bottom end, on the other hand,
the forecasters partition the total information and have zero overlap, i.e., $\delta =
\delta'/2$ and $\lambda = 0.0$.  Given that
moving down along these curves simultaneously increases information diversity and
$x_0$, both information diversity and the total amount of
information used by the forecasters are important yet separate determinants of  extremizing. This observation can guide practitioners towards proper extremization because  many  application specific aspects are linked to these two determinants. For instance, extremization can be
expected to increase in the number of forecasters,
subject-matter expertise, and human diversity, but to decrease in
collaboration, sharing of resources, and problem difficulty.
 
\section{PROBABILITY AGGREGATION}
\label{aggregation}
%This section first derives the revealed aggregator $p''$ for the
%general Gaussian model. By further assuming
%exchangeability among the forecasts, the revealed aggregator can be
%applied to any single pool of probabilities. After proving some
%properties of this aggregator, it is
%tested on real-world forecasts of one-time events. This
%illustrates one approach to estimating the information structures, and also provides some empirical evidence in favor of the
%partial information model.
% We then apply 
%this to the symmetric case, for which the oracular aggregator 
%was discussed in Section~\ref{compound}.  The revealed aggregator, 
%unlike the oracular aggregator, can be applied in practice.  
%We prove an extremizing result for this aggregator.
\subsection{Revealed Aggregator for the Gaussian Model}
Recall the multivariate Gaussian distribution (\ref{Nforecasters}) and 
collect all $X_{B_i} =
\Phi^{-1}(p_i)\sqrt{1-\delta_i}$ into a column vector $\boldsymbol{X} = (X_{B_1}, X_{B_2}, \dots,
X_{B_N})'$. If ${\bf\Sigma}_{22}$ is a
coherent overlap structure and ${\bf\Sigma}_{22}^{-1}$ exists, 
%then $X_{S} | \boldsymbol{X} \sim \mathcal{N}\left(\bar{\mu}, \bar{\Sigma}\right)$, where $\bar{\mu} 
%  = {\bf \Sigma}_{12} {\bf\Sigma}_{22}^{-1} \boldsymbol{X}$ and $\bar{\Sigma} = 1 - {\bf\Sigma}_{12} {\bf\Sigma}_{22}^{-1} {\bf\Sigma}_{21}$. 
%
%\begin{align*}
%\bar{\mu} &= \mu_1 + \Sigma_{12} {\bf\Sigma}_{22}^{-1} 
%  (\boldsymbol{X} - \boldsymbol{\mu}_2) 
%  = \Sigma_{12} {\bf\Sigma}_{22}^{-1} \boldsymbol{X} \\
% \bar{\Sigma}&= \Sigma_{11} - \Sigma_{12} {\bf\Sigma}_{22}^{-1} \Sigma_{21} 
% = 1 - \Sigma_{12} {\bf\Sigma}_{22}^{-1} \Sigma_{21}   \, 
%\end{align*}
%These expressions follow directly from the formulas of
%the conditional multivariate Gaussian distribution (see, 
%e.g., \citealt{ravishanker2001first}). 
%\citealt[Result~5.2.10, page~156]{ravishanker2001first}). 
then the revealed aggregator under the Gaussian model is
\begin{align}
p'' & =  \P\left(A  | \F''\right) =  \P\left(X_{S} > 0 | \boldsymbol{X}\right) = \Phi\left( \frac{{\bf\Sigma}_{12} {\bf\Sigma}_{22}^{-1} \boldsymbol{X}}
   {\sqrt{1 - {\bf\Sigma}_{12} {\bf\Sigma}_{22}^{-1} {\bf\Sigma}_{21}}}\right).
\label{GeneralAggregator} \,
\end{align}
%Furthermore, if $\one_N$ is a column vector of ones and 
%$\boldsymbol{P} = (P_{B_1}, P_{B_2}, \dots, P_{B_N})'$, 
%then the extremization parameter for $p''$ with respect to
%$\probit$ is given by 
%\begin{align*}
%\alpha  &= \frac{N \Sigma_{12} {\bf\Sigma}_{22}^{-1} 
%  \boldsymbol{X}}{\left(\boldsymbol{1}_N' \boldsymbol{P} \right) 
%  \sqrt{1 - \Sigma_{12} {\bf\Sigma}_{22}^{-1} \Sigma_{21}}}  \, .
%\end{align*}
% where $\tilde{p} = \P(A)$ is the prior probability discussed in Section \ref{prelim} and 
Applying (\ref{GeneralAggregator}) in practice requires an estimate of ${\bf\Sigma}_{22}$. If the forecasters make predictions about multiple events, it may be possible to model the different prediction tasks with a hierarchical structure and estimate a fully general form of ${\bf\Sigma}_{22}$. This can be formulated as a constrained (semi-definite) optimization problem, which, as was mentioned in Section \ref{prelim}, is left for subsequent work. Such estimation, however, requires the results of a large multi-prediction experiment which may not always be possible in practice. Often only a single prediction per forecaster is available. Consequently, accurate estimation of the fully general information structure becomes difficult. This motivates the development of aggregation techniques for a single event. Under the Gaussian model, a standard approach is to assume a covariance structure that involves fewer parameters. 
%As  was discussed earlier in Remark (\ref{item:specific}), the information structure can be estimated in one of three ways: by
%assumption, by estimation, or in a Bayesian manner.
The next subsection discusses a natural and non-informative choice.

%
%The next subsection analyzes the revealed aggregator under this simplification.
%\marginpar{motivate the choice of the model; contrast with fully general version}

%\section{Version 1}
\subsection{Symmetric Information}
\label{compound2}
%In this section we assume symmetry, meaning that ${\bf X}$ and
%$\Sigma$ are given by~\eqref{eq:symmetric}.  This assumption
%on the parameters of the model corresponds to
This subsection assumes a type of exchangeability among the forecasters.
While this is somewhat idealized, it is a reasonable choice in a
low-information environment where there is no historical or
self-report data to distinguish the forecasters.  The averaging
aggregators described in Section~\ref{sec:prior}, for instance, are
symmetric. Therefore, to the extent that they reflect an underlying
model, the model assumes exchangeability.  Under the Gaussian model, exchangeability suggests the compound
symmetric information structure discussed in Section \ref{compound}. This structure holds if, for example, the
forecasters use information sources sampled from a common distribution. 
The resulting revealed aggregator takes the form
\begin{align}
p_{cs}''
  &=\Phi\left(\frac{\frac{1}{(N-1)\lambda +1} 
  \sum_{i=1}^N X_{B_i} }{\sqrt{1- \frac{N\delta}{(N-1)\lambda +1} }}  
  \right), \label{CompoundAggre}
\end{align}
%where $\delta$ is the amount of information known by a forecaster, 
%$\lambda$ is the shared proportion of this information, and
 where $X_{B_i} =
\Phi^{-1}(p_i)\sqrt{1-\delta}$ for all $i = 1, \dots, N$. 
%Recall from section \ref{compound} that $\delta \in [0,1]$ can be 
%interpreted as the average amount of information known by an forecaster, 
%and $\lambda$ is the average proportion of the known information shared 
%between any two forecasters. 
%The domain restriction (\ref{rhoDomain}) on the parameters $\delta$ and $\lambda$ ensures that the term under
%the square-root in (\ref{CompoundAggre}) is always non-negative. 
%Unfortunately, this version is not as good as the oracular
%aggregator; the former is in fact a conditional expectation of the latter.

Given these interpretations, it may at first seem surprising that the
values of $\delta$ and $\lambda$ can be estimated in practice.
Intuitively, the estimation relies on two key aspects of the model: a) a better-informed forecast is likely to be further away from  the non-informative prior (see Figure \ref{marginals}); and b) two forecasters with high information overlap are likely to report very similar predictions. This provides enough leverage to estimate the
information structure via the maximum likelihood method.  Complete
 details for this are provided in 
Appendix B of the Supplementary Material.  Besides exchangeability, $p_{cs}''$ is based on very different modeling assumptions
than the averaging aggregators. The following proposition summarizes some of
its key properties.

\begin{proposition} \label{positiveThm}
%\phantom{Under symmetric information,}
\begin{enumerate}
\item[$(i)$] The probit extremization ratio between $p_{cs}''$ and $\probit$ is given by the
non-random quantity $\alpha(p_{cs}'', \probit) =  \gamma \sqrt{1 - \delta}/\sqrt{1-\delta\gamma}$, where $\gamma = N/((N-1)\lambda +1)$,
\item[$(ii)$] $p_{cs}''$ extremizes $\probit$ as long as $p_i \neq p_j$ for some $i \neq j$,
and
\item[$(iii)$] $p_{cs}''$ can leave the convex hull
of the individual probability forecasts.
\end{enumerate}
\end{proposition}
Proposition \ref{positiveThm} suggests that
$p_{cs}''$ is appropriate for combining probability
forecasts of a single event. This is illustrated on real-world forecasts in the next subsection. The goal is not to perform a thorough data analysis or model evaluation, but to demonstrate $p_{cs}''$ on a simple example.

\subsection{Real-World Forecasting Data}
\label{realData}
%\marginpar{edit this section}

Probability aggregation appears in many facets of real-world applications, 
including weather forecasting, medical diagnosis, estimation of credit default, and sports betting. 
%including weather forecasting  \citep{murphy1977reliability}, medical diagnosis  \citep{pepe2003statistical}, estimation of credit default
%\citep{kramer2006evaluating}, and sports betting \citep{dowie1976efficiency}.
 This section, however, focuses on predicting global events that are of particular interest to the Intelligence Advanced Research Projects Activity (IARPA). Since 2011, IARPA has posed about 100-150 question per year as a part of its ACE forecasting tournament. Among the participating teams, the Good Judgment Project (GJP) (\citealt{ungar2012good, mellers2014psychological}) has emerged as the clear winner. The GJP has recruited
thousands of forecasters to estimate 
probabilities of the events specified by IARPA.  The forecasters are told that their predictions
  are assessed using the Brier score (see Section \ref{BiasNoise}).  
In addition to receiving \$150 for
meeting minimum participation requirements that do not depend on
prediction accuracy, the forecasters receive status rewards for good
performance via leader-boards displaying Brier scores for the top 20
forecasters. Every year the top 1\% of the forecasters are
selected to the elite group of ``super-forecasters''. 
Note that, depending on the details of the reward structure, such a competition for rank may  eliminate the truth-revelation property of proper scoring rules (see, e.g., \citealt{lichtendahl2007probability}).
%The
%super-forecasters work in groups to make highly accurate predictions
%on the same events as the rest of the forecasters.

%\marginpar{this is not a case-study or through analysis of the data but an illustration of the framework}
% 

This subsection focuses on the super-forecasters in the second year of the tournament. Given that these forecasters were
elected to the group of super-forecasters based on the first year, 
%Our evaluation, however, only uses forecasts that were made
%during the second year. 
 their forecasts are
likely, but not guaranteed, to be relatively good. The group involves 44 super-forecasters collectively making predictions about  123 events, of which 23 occurred. 
 For instance, some of
the questions were: ``Will France withdraw at least 500 troops from Mali before 10 April 2013?", and ``Will a banking union be approved in the EU council before 1 March 2013?". 
%(see the Appendix in \citealt{satopaa} for more descriptions).  
%The forecasters were allowed to update their predictions as long as the
%questions remained active. 
%Some questions were active longer than the
%others. More specifically, the number of active days ranged from 3 to 284 days,
%with a mean of 96 days. The super-forecasters tended to update their predictions quite frequently. 
%The number of active days ranged from 3 to 284 days, with a mean of 96 days. 
%In this paper, however, only the most recent prediction made within the first three days of each problem is considered.  
%In the resulting
%dataset,
Not every super-forecaster made predictions about every event. In fact, the number of forecasts per event ranged from 17 to 34
forecasts, with a mean of 24.2 forecasts. To avoid infinite log-odds
and probit scores, extreme forecasts $p_i = 0$ and $1$ were
censored to $p_i = 0.001$ and $0.999$, respectively.

In this section aggregation is performed one event at a time
without assuming any other information besides the probability
forecasts themselves.  This way any performance improvements reflect better fit of the
underlying model and the aggregator's relative advantage in forecasting a single event. Aggregation accuracy is measured with the mean Brier score (BS): Consider $K$
events and collect all $N_k$ probability forecasts for event $A_k$ into a vector $\boldsymbol{p}_k \in [0,1]^{N_k}$. Then, BS for aggregator $g:[0,1]^{N_k} \to [0,1]$ is 
 \begin{align*}
\text{BS} &= \frac{1}{K} \sum_{k=1}^K \left(g(\boldsymbol{p}_k) - \one_{A_k}\right)^2.
 \end{align*}
This score is defined on the unit interval with lower values indicating higher accuracy. For a more detailed performance analysis, it decomposes into three additive components: reliability (REL),
resolution (RES), and uncertainty (UNC). This assumes
that the aggregate forecast $g(\boldsymbol{p}_k)$ for all $k$ can only take discrete values $f_j \in
[0,1]$ with $j = 1, \dots, J$. Let $n_j$ be the number of times $f_j$
occurs, and denote the empirical frequency of the corresponding events
with $o_j$.  Let $\bar{o}$ be the overall empirical frequency of
occurrence, i.e., $\bar{o} = \frac{1}{K} \sum_{k=1}^K  \one_{A_k}$. Then,
 \begin{align*}
\text{BS} &= \text{REL} - \text{RES} + \text{UNC}\\
&= \frac{1}{K} \sum_{j=1}^J n_j (f_j - o_j)^2 - \frac{1}{K} \sum_{j=1}^J n_j (o_j - \bar{o})^2 + \bar{o}(1-\bar{o}).
 \end{align*}
  In this decomposition low REL represents good calibration. If a calibrated aggregate is also confident, it exhibits high RES.  Therefore the combination of good calibration and high confidence leads to low BS. The corresponding forecasts are likely to be very close to $0$ and $1$, which is more useful to the decision-maker than the naive forecast $\bar{o}$. The final term UNC equals the BS for $\bar{o}$ and hence provides a reference point for interpreting the performance of the aggregator. 
%In fact, any performance improvements can be interpreted relative to the final term UNC that equals the BS for $\bar{o}$.
% and serves as a benchmark for interpreting 
%\marginpar{why do we care about the individual components; and why we care about RES and REL}

%subject to
%being calibrated is preferred \citep{Ranjan08}. The
%uncertainty term quantifies overall uncertainty among the events and
%does not depend on the aggregate forecasts.
 
% % latex table generated in R 3.0.3 by xtable 1.7-3 package
%% Tue Aug 26 11:03:33 2014
%\begin{table}[t]
%\centering
%\caption{The mean Brier scores (BS) with its three components, reliability (REL), resolution (RES), and uncertainty (UNC), for the aggregated super-forecasts.}
%\begin{tabular}{lcccc}
%  \hline\hline
%Aggregator & BS & REL & RES & UNC \\ 
%  \hline
%$\bar{p}$ & 0.1306 & 0.0351 & 0.0565 & 0.1520 \\ 
% $\plog$ & 0.1246 & 0.0262 & 0.0536 & 0.1520 \\ 
% $\probit$ & 0.1241 & 0.0272 & 0.0551 & 0.1520 \\ 
% $p''$ & 0.1199 & 0.0198 & 0.0520 & 0.1520 \\ 
%   \hline
%\end{tabular}
%\label{BrierTable}
%\end{table}

 % latex table generated in R 3.0.3 by xtable 1.7-3 package
% Tue Aug 26 11:03:33 2014
\begin{table}[t]
\centering
\caption{The Mean Brier Scores (BS) with Its Three Components, Reliability (REL), Resolution (RES), and Uncertainty (UNC), for Different Aggregators.}
\begin{tabular}{lcccc}
  \hline\hline
Aggregator & BS & REL & RES & UNC \\ 
  \hline
$\bar{p}$ & 0.132 & 0.026 & 0.045 & 0.152 \\ 
 $\plog$ & 0.128 & 0.025 & 0.048 & 0.152 \\ 
 $\probit$ & 0.128 & 0.023 & 0.047 & 0.152 \\ 
 $p_{cs}''$ & 0.123 & 0.020 & 0.049 & 0.152 \\ 
   \hline
\end{tabular}
\label{BrierTable}
\end{table}

Table \ref{BrierTable} presents results for $\bar{p}$, $\plog$, $\probit$, and $p_{cs}''$ under the super-forecaster data. Empirical approaches were not considered for two
reasons: a) they do not reflect an actual model of forecasts; and b)
they require a training set with known outcomes and hence cannot be
applied to a single event. 
Overall, $\bar{p}$ presents the worst performance. Given that  $\probit$ and $\plog$ are
very similar, it is not surprising that they have almost identical scores. The revealed aggregator $p_{cs}''$ is both the most resolved and calibrated, thus achieving the lowest BS among all the aggregators. This is certainly an encouraging
result. It is important to note that $p_{cs}''$ is only the first attempt at partial
information aggregation. 
%The framework points a clear direction for continued improvement. In particular, 
More elaborate information structures and
estimation procedures, such as shrinkage estimators, are very likely to lead to many further
improvements.

% \textcolor{red}{Mention that unlike the other approaches points the way for improved development.}

\section{SUMMARY AND DISCUSSION}
\label{discussion}
This paper introduced a probability model for predictions made by
a group of forecasters.  The model allows for interpretation of some
of the existing work on forecast aggregation and also clarifies
empirical approaches such as the {\em ad hoc} practice of
extremization.  The general model is more plausible on the micro-level
than any other model has been to date. Under this model, 
some general results were provided. For instance, the
\textit{oracular} aggregate, which uses all the forecasters' information (Proposition \ref{condInd}), is more likely to be
more extreme than one of the common benchmark aggregates, namely 
$\probit$ (Proposition~\ref{positiveProbThm}).  Even though
no real world aggregator has access to all the information of the
oracle, this result explains why extremization is almost certainly
called for.  More detailed analyses  were performed under several
specific model specifications such as zero and complete information
overlap (Section~\ref{disjoint}), and fully symmetric information (Section~\ref{compound}).  Even though the zero and complete
information overlap models are not realistic, except under a very
narrow set of circumstances, they form logical extremes that illustrate the main drivers of good aggregation. The symmetric model is somewhat
more realistic. It depends only on two parameters and therefore allows us to visualize the effect of model parameters on the optimal amount of
extremization (Figure~\ref{LevelplotsOracle}).  Finally, the
{\em revealed} aggregator, which is the best in-practice aggregation
under the partial information model, was discussed. The discussion provided a general formula for
this aggregator (Equation~\ref{GeneralAggregator}) as well as its
specific formula under symmetric information
(Equation~\ref{CompoundAggre}). The specific form was applied to real-world forecasts 
of one-time events and shown to 
outperform other model-based aggregators.

It is interesting to relate our discussion to the many empirical
studies conducted by the Good Judgment Project (GJP) (see Section
\ref{realData}).  Generally
extremizing has been found to improve the average aggregates
\citep{mellers2014psychological, satopaa, satopaa2014probability}.
The average forecast of a team of super-forecasters, however, often
requires very little or no extremizing.  This can be explained as follows.  The super-forecasters are
highly knowledgeable (high $\delta$) individuals who work
 in groups (high $\rho$ and $\lambda$).  Therefore, in
Figure \ref{LevelplotsOracle} they are situated around the upper-right
corners where almost no extremizing is required.  In other words,
there is very little left-over information that is not already used in each
forecast.  Their forecasts are highly convergent and are likely to be
already very near the oracular forecast.  The GJP forecast data also
includes self-assessments of expertise.  Not surprisingly, the greater
the self-assessed expertise, the less extremizing appears to have been
required. This is consistent with our interpretation that high values
of $\delta$ and $\lambda$ suggest lower extremization.

The partial information framework offers many directions for future research.  One involves estimation of parameters.  
%The Gaussian model together with good procedures for estimating
%the information overlap among two or more forecasters is likely to lead to many
%improvements in probability aggregation. 
%Given a stream of forecasts
%of reasonable length, it is possible to estimate $|B_i|$ and $|B_i
%\cap B_j|$ with a maximum likelihood procedure akin to the details
%described in the Appendix. 
In principle, $|B_i|$ can be
estimated from the distribution of a reasonably long probability stream. Similarly, $|B_i \cap B_j|$ can be estimated from the correlation of the two parallel streams.  Estimation of higher order intersections, however, seems more
dubious. In some cases the higher order intersections have been found to be irrelevant to the aggregation procedure. For instance, \citet{degroot1991optimal} show that it is enough to consider only the pairwise conditional (on the truth) distributions of the forecasts when computing the optimal weights for a linear opinion pool. 
%Computations with $N=3$ show that, at least in some cases,
%higher order intersections are not relevant to the revealed
%aggregator. 
Theoretical results on the significance or insignificance
of higher order intersections under the partial information framework
would be desirable. 

%Unfortunately, the Gaussian model cannot incorporate intersections beyond the second order. Therefore, if higher order intersections turn out to be relevant for aggregation, it may be necessary to develop a different, more complex partial information model. 

Another promising avenue is the Bayesian approach. In many applications with small or moderately sized datasets, Bayesian methods have been found to be superior to the likelihood-based alternatives. Therefore, given that the number of forecasts on a single event is typically quite small, a Bayesian approach is likely to improve the predictions of one-time events. Currently, we have
work in progress analyzing a Bayesian model but there are many, many
reasonable priors on the information structures. 
%Generally, two forecasters
%with very similar or even identical predictions are more likely
%to have very similar information sets. 
% In a Bayesian model, one would
%expect more extremization from the probit opinion pool when the set of
%forecasts is varied than when it is relatively homogeneous.
 This avenue should
certainly be pursued further, and the results tested against other high
performing aggregators.

\section{SUPPLEMENTARY MATERIALS}
\label{supplementary}
\begin{description}
\item[Technical Details:] This supplementary material includes a) proofs for Propositions \ref{CorrelationPolytope}, \ref{condInd}, \ref{positiveProbThm}, and \ref{positiveThm}; b) derivation of Equation \ref{feasible}; and c) instructions on how to estimate the model parameters under symmetric information. (.pdf file)
\end{description}

\bibliographystyle{apalike}
\bibliography{biblio}		% expects file ''myrefs.bib''

\end{document}